\documentclass[fleqn,10pt]{wlscirep}
\usepackage{bbold}
\title{Artificial Life in Quantum Technologies}

\author[1,*]{Unai Alvarez-Rodriguez}
\author[1]{Mikel Sanz}
\author[1]{Lucas Lamata}
\author[1,2]{Enrique Solano}
\affil[1]{Department of Physical Chemistry, University of the Basque Country UPV/EHU, Apartado 644, 48080 Bilbao, Spain}
\affil[2]{IKERBASQUE, Basque Foundation for Science, Maria Diaz de Haro 3, 48013 Bilbao, Spain}

\affil[*]{unaialvarezr@gmail.com}

\begin{abstract}
We develop a quantum information protocol that models the biological behaviours of individuals living in a natural selection scenario. The artificially engineered evolution of the quantum living units shows the fundamental features of life in a common environment, such as self-replication, mutation, interaction of individuals, and death. We propose how to mimic these bio-inspired features in a quantum-mechanical formalism, which allows for an experimental implementation achievable with current quantum platforms. This study paves the way for the realization of artificial life and embodied evolution with quantum technologies.
\end{abstract}

\begin{document}

\flushbottom
\maketitle
\thispagestyle{empty}

\section*{Introduction}
In the last decades, the novel field of artificial life has enabled researchers to recreate biological behaviours with controllable inanimate platforms in the laboratory~\cite{bed03}. Its goals are diverse, ranging from the comprehension of the emergence of life to the explanation of the appearance of dynamical hierarchies that give rise to complexity. Examples of the latter are consciousness at the single agent level or social organization at the group level. Self-replication and self-organization have already been achieved in this context based on fundamental interactions between the artificial living entities called~{\it individuals}~\cite{ag14}.  Moreover, techniques developed in artificial life have been applied into different research lines, e.g., by modeling the formation of biological tissues~\cite{la14}, and explaining the dynamical structure of fluids~\cite{flu}. In particular, software-based artificial life consists of computational algorithms of evolving individuals. This area has produced some prominent models like the Game of Life \cite{gard} or Tierra \cite{ray}, that in most cases were developed using classical techniques, with few examples in the quantum domain \cite{lif,aqgl,qgl}. 
 
It is known that certain quantum information protocols \cite{yoah, yiah} can be performed efficiently in terms of speed or number of resources. Therefore, it seems natural to look for the consequences of introducing quantum mechanics in artificial life models, and establish analogies and connections between these two seemingly unrelated fields. There are already some preliminary results in the realm of quantum evolution \cite{ma} and quantum learning agents \cite{hb}.  Here, we would like to focus on the concept of quantum biomimetics that was introduced in Ref.~\cite{yo} as mimicking macroscopic biological behaviours at the quantum microscopic level, for the sake of quantum information fundamentals and applications. 
  
In this article, in the context of quantum biomimetics, we propose a quantum model of artificial life that aims at reproducing fundamental biological behaviours with controllable quantum platforms. Along these lines, we define quantum {\it individuals} that can be born, evolve, interact, mutate, and die, while they propagate and decohere in a common environment. These concepts are designed to be implementable with current quantum technologies. Hence, we discuss an experimental realization in trapped ions, superconducting circuits, and integrated quantum photonics. Related ideas have already been echoed by the scientific community \cite{mit}. 

It is noteworthy to mention that the proposed quantum biomimetic ideas should be considered as a free creation of a possible quantum evolution model at the microscopic quantum level. In this sense, while related to standard fields as artificial intelligence, machine learning, cellular automata, artificial living systems, and the like, these ideas cannot be framed uniquely in any of them. 

\section*{Results}
\begin{figure}[h!!!]
\centering
\includegraphics[width=14cm]{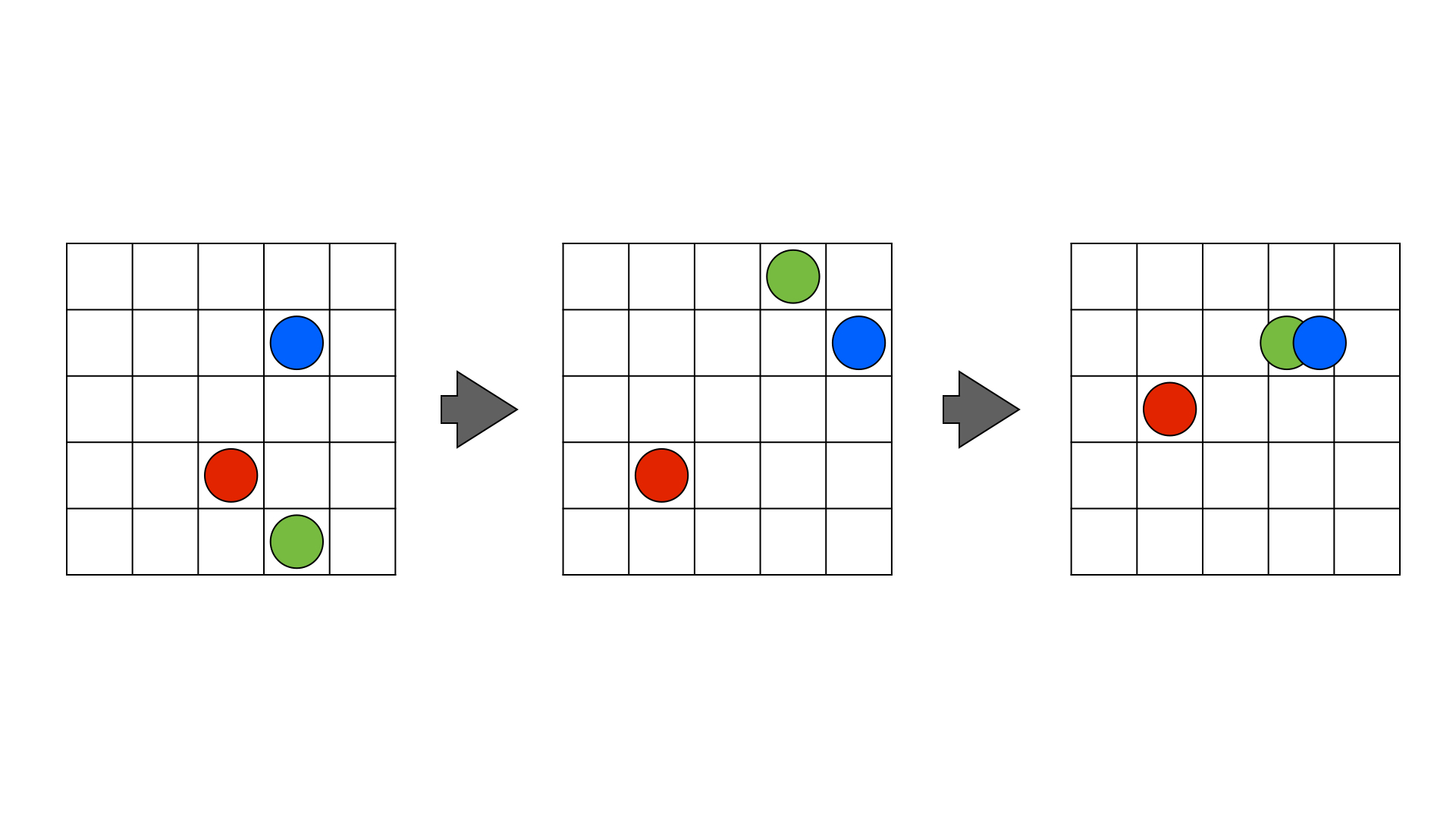}
\caption{{\bf Spatial Dynamics.} Coloured circles represent individuals that move in discrete time-steps along a periodic spatial grid. This schematically illustrates how two individuals can share the same location and interact.}
\label{grid}
\end{figure}

The individuals, which are our quantum artificial living units, are described by two-qubit states. Inspired in the biological mechanism of self replication and evolution, one of the qubits in a quantum living unit represents its {\it genotype} and the other qubit corresponds to its {\it phenotype}. In our model, the information with the characteristics of the living feature is codified in the genotype, and is transmitted through generations. Additionally, the genotype encodes the lifetime of the individual and its role in the trophic chain. On the other hand, the phenotype is the expression of the genotype under the influence of an environment. Specifically, the phenotype carries information about the age of the individual, which is encoded in the time elapsed in the evolution of the qubit from its initial state to the dark state of the environment. The information exchange between the genotype and the phenotype is produced at the creation of a newborn individual during the self-replication process. Moreover, the analogy with biological mutation is provided by adding the possibility of modifying the genetic qubit or introducing errors in the replication stage. Finally, the individuals live in a discrete spatial grid in which they move by virtue of a random process, see Fig \ref{grid}. When they share the same location, they interact with each other, in an operation that modifies the phenotype but preserves the genotype. In the following, we explain in detail the physical operations underlying the different aspects of our model.

\subsubsection*{Self-replication} The mechanism of self-replication is based on the cloning of partial quantum information explained in Ref. \cite{yo}. The qubit in which the information is encoded is coupled with an ancillary state that belongs to the environment. The information transmission consists in an entangling operation that distributes the information throughout the two-qubit quantum state. After the partial quantum cloning process, the information can be retrieved from both subspaces independently. Specifically, the expectation value of a desired operator $\theta$ in any quantum state $\rho_0$ can be propagated into the following generations $\rho_1$ with the use of an ancillary state $\rho_A$ and a unitary operation $U$. 
\begin{equation}
\label{bcqo}
\langle \theta \rangle_{\rho_{0}}=\langle \theta \otimes \mathbb{1} \rangle_{\rho_1}=\langle \mathbb{1}\otimes \theta \rangle_{\rho_1}, \qquad \rho_1=U(\rho_0 \otimes \rho_A)U^{\dag}
\end{equation}
In particular, we are using the expectation value of $\sigma_z$ as the genotype of the individuals and the $U_{\rm CNOT}$ gate as the cloning operation. Different combinations of observables and cloning operations would also satisfy the partial quantum cloning criteria, but we have selected $\sigma_z$ because it is diagonal in the basis given by the steady state of the environment. By construction, unlimited copies of the ancillary quantum state $\rho_A$ are available in our model everywhere in the spatial grid, and they belong to the dark state of the dynamics that governs the interaction between individuals and environment. A new individual is produced in two steps. In the first one, the genotype belonging to the procreator individual is copied onto an ancillary state in order to produce a new genotype. In the second step, the genotype of the new generation is copied onto another ancillary state in order to produce the phenotype. There is a fixed probability of self-replication that is equal for all individuals, since it does not depend on the genotype. The only requirement for the generation of a newborn individual via the self-replication process is that the procreator individual is alive. This property is encoded in the phenotype and depends on the interaction of the individual with the environment, as we describe below.
 
\subsubsection*{Environment} 
When the new individual is created, its genotype and phenotype exactly contain the same information. However, they progressively differentiate as the system evolves due to the coupling of the phenotype with the environment. This mechanism mimics a crucial feature in natural selection, namely, the preservation of the genetic information throughout successive generations. At the same time, the phenotype is degraded due to the interaction with the environment, which concludes with the death of the quantum living unit. 
 
The dissipation is modeled with a Lindblad master equation, whose steady state corresponds to the ancillary state of the copying process, $\rho_A$. We define the Lindblad operators as acting in the natural basis of the environment, given by the $\sigma_z$ basis of the self-replication process, i.e.,
\begin{equation}
\label{lind} 
\dot{\rho}=\mathcal{L}\rho=\gamma (\sigma \rho \sigma^\dag -\frac{1}{2}\sigma^\dag \sigma \rho -\frac{1}{2} \rho \sigma^\dag \sigma ), \qquad  \sigma=|0\rangle\langle1|, \qquad \rho_A=|0\rangle\langle0|.
\end{equation}
By evolving the system under this Lindbladian $\mathcal{L}$, all individuals end up in the state $\rho_A$. Therefore, we use this physical register to simulate the death of the quantum living unit. The cycle closes since the dead individuals serve as ancillary states for the new generations.
 
We can illustrate the processes of self-replication and aging by dissipation with a generic example in which an individual and its progeny are created out of a precursor genotype. Let us suppose that, initially, there is a single genetic qubit $\rho_g$ copied into a phenotype qubit in order to create an individual $\rho_0$,
 \begin{eqnarray}
&& \rho_g=\left( \begin{array}{cc} a&b-ic\\b+ic&1-a \end{array} \right), \qquad \rho_0 (t=0)=U (\rho_g \otimes \rho_A) U^\dag, 
\end{eqnarray}
The individual evolves under the dissipative dynamics with $\sigma=\mathbb{1}\otimes|0\rangle\langle1|$.
\begin{eqnarray}
&& \rho_0(t)=\left( \begin{array}{cccc} a&0&0&(b-ic)e^{-\frac{1}{2} \gamma t}\\0&0&0&0\\0&0&(1-a)(1-e^{-\gamma t})&0\\(b+ic)e^{-\frac{1}{2} \gamma t}&0&0&(1-a)e^{-\gamma t}  \end{array}\right)
 \end{eqnarray}
The expectation value of $\sigma_z$ remains constant in the genotype subspace, but exponentially decays in the phenotype subspace,
\begin{eqnarray}
\langle \sigma_z \rangle_g=2a-1, \qquad \langle \sigma_z\rangle_p=1-2 e^{-\gamma t} (1-a). \end{eqnarray}  
The expectation value $\langle \sigma_z\rangle_p(t)$ measuring the age of the individual depends on a single genetic parameter $a$, and on the elapsed time in the evolution from the birth of the quantum living unit until its death. The death age $t_d$ of the individual is achieved when $\langle \sigma_z \rangle_p (t_d)=1-\epsilon$, for fixed $\epsilon$. 
  
\begin{figure}[h]
\centering
\includegraphics[width=18cm]{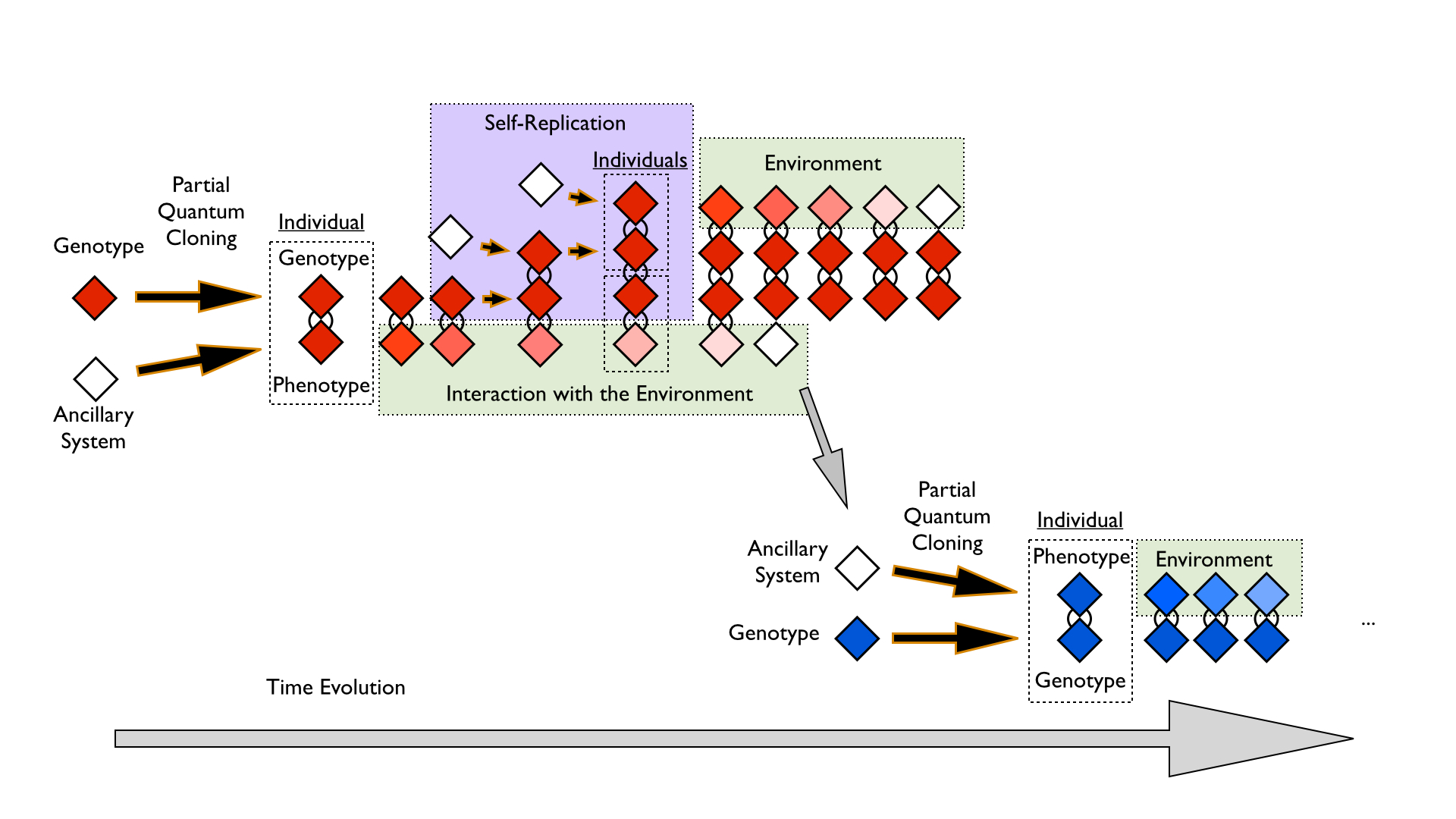}
\caption{{\bf Self-replicating Individuals.} We schematize a timeline in the evolution of individuals. Firstly, an individual is created out of a primordial genotype by combining it with an ancillary system. Then, the individual interacts with the environment which produces a gradual loss in the information of the phenotype, until the death of the individual. In the mean time, the living unit generates a new individual via the self-replication process. Additionally, the remaining phenotype can be used as a new ancillary state for the creation of another individual.}
\label{fg}
\end{figure}

The individual continues its evolution until the self-replication protocol begins, as described in Fig.~\ref{fg}. This protocol consists in coupling an ancillary state $\rho_A$ with the individual $\rho_0$ and performing the $U$ operation between the genotypes of the two individuals. The next step is to use another ancillary state from the environment and perform the $U$ gate between the genotype and phenotype of the new individual. We denote the self-replication time by $t_1$, after which the whole system is coupled with the environment via the Lindblad master equation for a time $t_2$. The expectation values of $\sigma_z$ for the genotype and phenotype of the two individuals are given by, 
\begin{equation}
\langle \sigma_z \rangle_{g1}=2 a-1, \qquad \langle \sigma_z \rangle_{p1}=1-2 e^{-\gamma (t_1+t_2)}(1-a), \qquad \langle \sigma_z \rangle_{g2}=2a-1, \qquad \langle \sigma_z \rangle_{p2}=1-2 e^{-\gamma t_2}(1-a).
\end{equation}

\subsubsection*{Mutations} The mutation process enhances diversity in biological systems, which is a fundamental property of Darwinian evolution. The system adaptability to a changing environment is closely related to its mutation capacity. In our model, the mutation is a physical operation upon the individuals, which changes their genotype. We distinguish between two types of mutations,  implemented with a small probability. On the one hand, there are spontaneous mutations upon the genotype subspace of an individual, given by the unitary matrix $M$,
\begin{equation}
\label{mut1}
M=\left( \begin{array}{cc} \cos\theta&\sin\theta \\ \sin\theta&-\cos\theta \end{array} \right).
\end{equation}
The mutation parameter $\theta$ is random and different for every mutation event in order to maximize the Hilbert space region spanned by the quantum states, and therefore maximize the biological diversity. On the other hand, there are mutations associated with errors in the copying process, modeled by imperfect cloning unitary operations $U_M$,
\begin{equation}
\label{mut2}
U_M (\theta)=\mathbb{1}_4 +\frac{1}{2}\left(\begin{array}{cc} 0&0\\0&1 \end{array}\right) \otimes  \left(\begin{array}{cc} -1&1\\1&-1 \end{array}\right) (\cos\theta+i\sin\theta+1).
\end{equation}
 The difference between these mutation operations relies on the fact that $M$ does not affect the phenotype of the mutated individual, while $U_M$ changes both the genetic information and the lifetime.
 
\subsubsection*{Interactions} In the formalism explained so far, the natural selection mechanism is completely biased towards the long-living individuals, corresponding to $a\sim0$ and $\langle \sigma_z \rangle\sim-1$. As the self-replicating probability is equal for all of them, the long-living individuals dominate the system due to the generation of a larger offspring. On the contrary, the possibility of interactions between pairs of individuals favors the short living individuals, and restores the equilibrium between long and short life genotypes. The idea behind the process is that individuals conditionally interchange their genotype when meeting each other. The conditionality depends on the genotype, namely, the interaction is minimal for equal genotypes $a_1=a_2$ and maximal for opposite ones, $a_1=1$, $a_2=0$ and viceversa. The physical operation $U_I$ that performs the interaction between two individuals is
\begin{eqnarray}
\label{inter}
U_I & = &  k_1\otimes (\mathbb{1}_2\otimes k_1 \otimes \mathbb{1}_2 + k_1 \otimes k_4 \otimes k_1+k_4 \otimes k_4 \otimes k_4 + k_2 \otimes k_4 \otimes k_3 +k_3 \otimes k_4 \otimes k_2) \nonumber \\  && +  k_4 \otimes (\mathbb{1}_2\otimes k_4 \otimes \mathbb{1}_2 + k_1 \otimes k_1 \otimes k_1+k_4 \otimes k_1 \otimes k_4 + k_2 \otimes k_1 \otimes k_3 +k_3 \otimes k_1 \otimes k_2),
\end{eqnarray}
with $k_1=|0\rangle\langle0|, k_2=|0\rangle\langle1|, k_3=|1\rangle\langle0|, k_4=|1\rangle\langle1|.$ Here, $U_I$ is a logical gate acting on two control qubits and two target qubits, which is defined in the computational basis as follows. When the control qubits are equal, the target qubits remain unchanged. On the contrary, when the control qubits are different, the target qubits are exchanged. In particular, we have selected the first and the third subspaces as control qubits. A direct implication of the interaction process is that, when a short-living individual encounters a long-living one, the lifetime of the former increases, while the lifetime of the latter decreases. Therefore, to have a genotype parameter $a\sim1$ is more advantageous in this case, because it corresponds to the role of a predator in our model.  According to this, every individual is a combination of predator and prey depending on its instantaneous phenotype and the local environment. Notice that the role of predator and prey can be completely interchanged in consecutive events for individuals that have previously interacted. The probability of this second-order event is low, and it depends on the interaction rate, which at the same time depends on the spatial distribution of the individuals and the grid geometry. Therefore, for each initial state of a particular individual, there are a grid geometry and distribution of the other quantum living units, which respectively optimize the survival probability and the interaction rate for the system of individuals.

\subsubsection*{Spatial Dynamics}
\begin{figure}[h]
\centering
\includegraphics[width=13cm, trim=0cm 0cm 0cm 0cm]{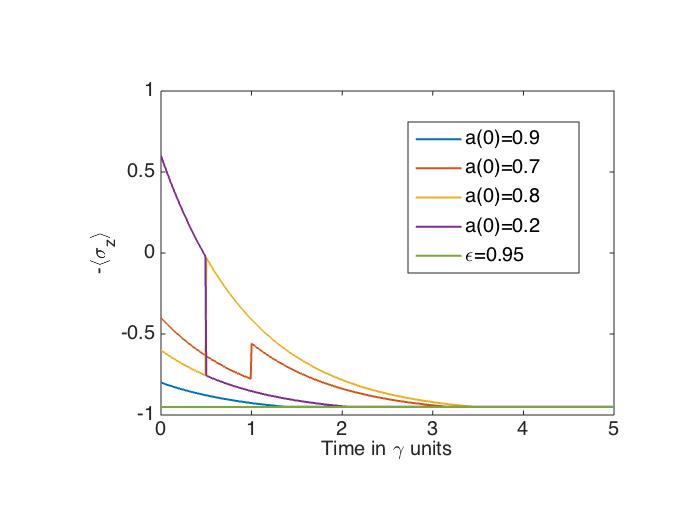}
\caption{{\bf Processes in the artificial life model.} We plot a combination of the genotype and phenotype of the individuals that depends on the genotype parameter $a$ and the time $t$ in which the individual has been in contact with the environment. The life quantifier is $\langle\sigma_z\rangle_p$, whose asymptotic value represents the death of the individual in the bottom of the figure. Each colour is associated with a basic process in the model: the blue line $(a=0.9)$ shows the dissipation of the phenotype due to the environment, the orange line $(a=0.7)$ depicts a mutation process event, and the yellow and purple lines, $a_1=0.8$ and $a_2=0.2$ respectively, illustrate an interaction process.}
\label{dos}
\end{figure}

In our model, the individuals live in a two-dimensional grid divided into cells, as seen in Fig.~\ref{grid}. The spatial distribution of the individuals determines the interaction rate, because they only interact when occupying the same cell. The displacement along the grid is a random process, and the proposed model allows for two or more individuals occupying the same cell. Furthermore, the grid can be split into distinct spatial regions with different properties: mutation rate, self-replication probability, and coupling constant with the environment. These are basic properties of the model associated with physical processes encoded in the genotype and phenotype, as we show in Fig.~\ref{dos}.

\subsection*{Numerical Simulations} In this subsection, we explain the dynamical numerical simulations based on our model. Classical artificial life models may address relevant questions about the properties of the self-replicating units: Which kind of individuals have maximized their survival probability by adapting to the environmental characteristics and the presence of other individuals? Is this an asymptotic behaviour or on the contrary is a part of a cycle in populations? In the case in which more than one species dominates how distinct are their genotypes, and has any complex spatial organization emerged? In our case, the time evolution is computationally hard due to the exponential growth of the Hilbert space dimension with the number of individuals.  Therefore, answering these questions in our context may motivate the realization of our proposal in an experiment on a controllable quantum platform, as discussed below. In our current analysis, we numerically examine other interesting properties, such as the information spreading and the genotype and phenotype diversities. 

\begin{figure}[h!!]
\centering
\includegraphics[width=17cm, trim=0cm 0cm 0cm 0cm]{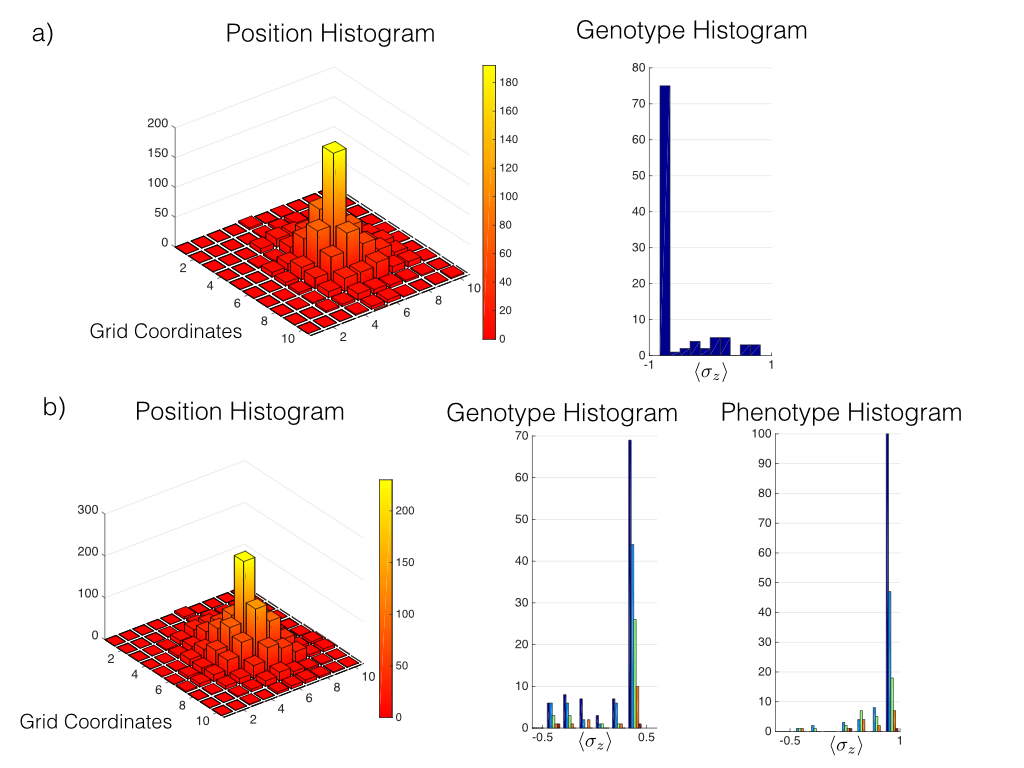}
\caption{{\bf Numerical simulation of replicating and non-replicating individuals.} These plots are obtained evolving an individual with a random genotype at a given initial position during a time $\gamma t =10$. The position histogram shows the accumulation of the paths covered by the living unit for all runs of the simulation. The genotype histogram shows the expectation value of $\sigma_z$ in the genotype subspace after each simulation. The initial individual does not self-replicate in ($a$), while it does self-replicate in ($b$), therefore the secondary peaks in the genotype and phenotype histograms appear in ($b$) as a consequence of the newborn individuals.}
\label{one}
\end{figure}

In order to study the information spreading and the mean path of the individuals, we generate a position histogram of the quantum living units for a large number of realizations, with fixed initial conditions. Moreover, the density in each cell is related to the probability of finding a single individual in that particular position. Therefore, we can estimate the interaction and self-replication rates by comparing position histograms corresponding to different parameters in the model, as shown in Fig. \ref{one}. Larger values of the position histogram peaks indicate the presence of additional individuals, and therefore, a larger amount of self-replicating events. 

\begin{figure}[h!!]
\centering
\includegraphics[width=17cm, trim=0cm 0cm 0cm 0cm]{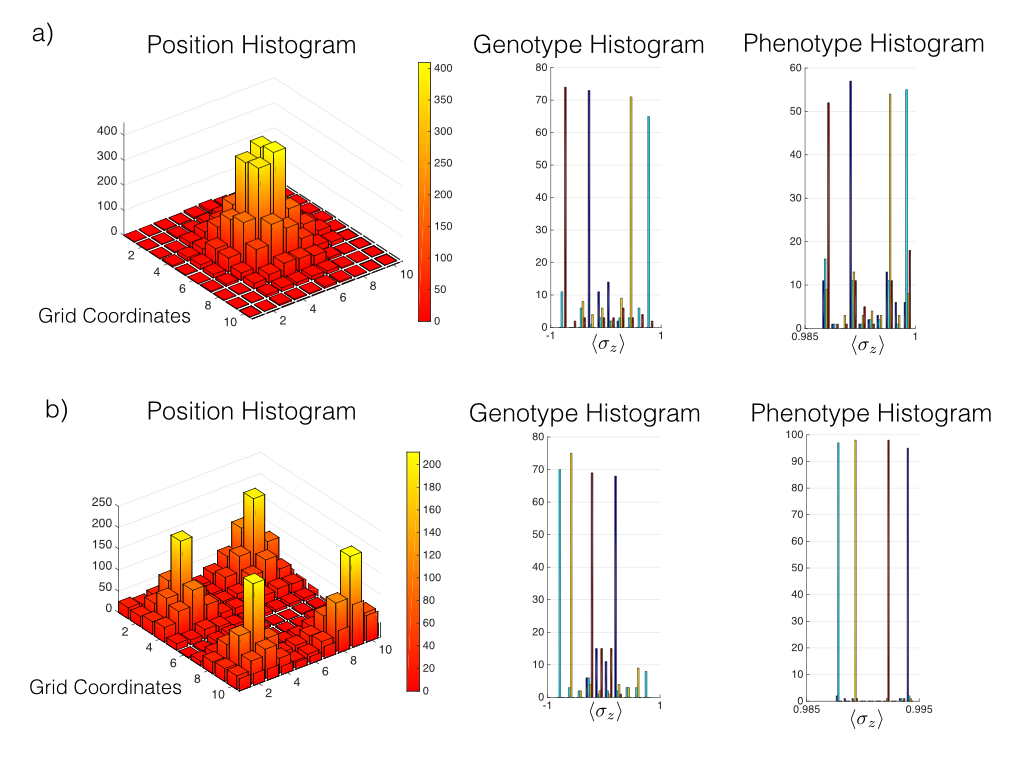}
\caption{{\bf Numerical simulation of interacting and non-interacting individuals.} We have limited the simulation of our model to $4$ initial individuals until a maximum time of $\gamma t=10$. No self-replicating events are allowed, and therefore, the diversity in the phenotype is due to interaction events. ($a$) The interaction rate is increased because of the short distances among the living units. ($b$) The interaction rate is small because of the long distances among the living units. Therefore, the secondary peaks in the phenotype histogram ($a$) are associated with interaction events which exchange the phenotype of individuals. }
\label{leka}
\end{figure}

Additionally, we have produced genotype and phenotype histograms that show the expectation value of $\sigma_z$ in each subspace of every individual, providing a snapshot of the final state of the simulation. These histograms evidence the preservation of the initial genotypes, the decay of the phenotypes, as well as the mutations and interactions which give origin to diversity, see Fig.~\ref{leka}. The deviation of the initial information in the genotype is related to mutation events. In contrast, the deviation from the initial peak in the phenotype histogram is related to interaction events, when simultaneously produced in two individuals. Otherwise, the change in the phenotype is due to a mutation followed by a self-replication event.

\begin{figure}[h!!]
\centering
\includegraphics[width=17cm, trim=0cm 0cm 0cm 0cm]{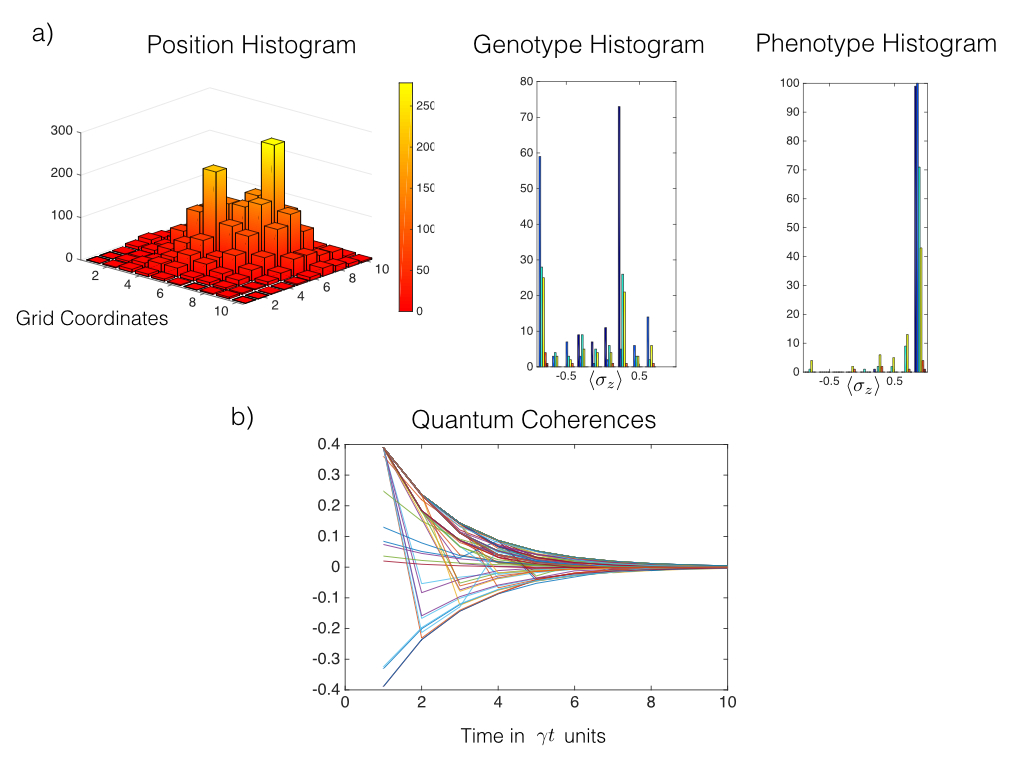}
\caption{{\bf Numerical simulation with quantum coherences.} Data compilation for 100 simulations of the time evolution of two initial individuals allowed to self-replicate. ($a$) The position histogram shows two peaks of the initial spatial distribution and the spread of the individuals as time increases. The effects of the copying process are illustrated in the genotype histogram with the presence of secondary peaks behind the principal one that corresponds to the original individual. The small peaks in the phenotype histogram represent the newborn individuals.  In ($b$), we have depicted the expectation value of $\sigma_x$ in all subspaces, $\langle \sigma_{x}^{\otimes 2 n} \rangle$, where $n$ is the number of individuals.  These quantum coherences give us information about the history of the individual. Therefore, we can infer the mutation and interaction events as well as the dissipative dynamics.}
\label{kale}
\end{figure}

\subsubsection*{Quantumness} The entanglement among different individuals allows us to clone the classical information and propagate the quantum coherences of the initial quantum living units to the successive generations~\cite{yo}. Nevertheless, the fact that our model requires the solution of a Lindblad master equation for a large number of entangled qubits, restricts the solution of the model to a small number of individuals and short times. Therefore, an experimental implementation of our model would enable to reach a larger number of individuals and longer times, hence, to increase the complexity of the system. Some of the features of our model are purely quantum, for instance, the entanglement among individuals with the same origin permits to measure collective correlations of the whole family. In this way, we can both distinguish between individuals with the same or different genotype, and individuals with the same or different origin. The physical mechanism in which we are basing our claim is the propagation of the collective expectation value $\langle\tau\rangle$ in \cite{yo}, see Fig.~\ref{kale}. Additionally, the individuals are in a superposition of a prey and predator, which allows us to simulate a trophic chain behaviour encoded in the two qubits that conform each individual.

\subsubsection*{Connections with other fields} In the same way that classical artificial life models can be applied in other areas of science, we think that our protocol is closely related with some aspects of quantum information theory. One can understand the model as a naturally emergent maximization problem of survival under the rules imposed by the environment, mutation rate, grid geometry and self-replication rate. The rules are external and tunable, which means that we could encode optimization problems and solve them by using an artificially-engineered natural-selection quantum mechanism.  Furthermore, our model of natural selection can be related to quantum game theory and quantum learning, if we consider that each of the genotypes encodes a strategy, and the environment together with the aforementioned elements establish the rules of the game.  Typically, the players with winning strategies survive, a fact that makes the rewarding mechanism an intrinsic part of the game. Under these analogies, it may also be possible to analyze the robustness of the optimization process or the strategy by changing the parameters describing the proposed model.

We point out that our quantum biomimetic model is not a particular case of a quantum cellular automaton (QCA). In a QCA, the information is encoded in the spatial grid, while in our model the information is stored in individuals that displace along the spatial grid. Therefore, the time evolution in the QCA system is by construction different to the time evolution in our model. In a QCA, the spatial lattice state at time $t$ is obtained by updating the one at $t-1$ according to the automaton transition function \cite{lif}.  

\subsection*{Proposal for an experimental implementation}
Our model may straightforwardly be implemented on a variety of quantum platforms, which would be justified due to the theoretical interest and the computational difficulty of classical simulations to answer several relevant questions. Here, we provide an encoding of the information in the respective qubits and the sequences of gates implementing our dynamics for trapped ions~\cite{ion,tis,npi}, superconducting circuits~\cite{scc,och}, and quantum photonics~\cite{pho,nmar,loqc}.  

The $U_{\rm CNOT}$ implements the self-replication process as explained in Eq. \eqref{bcqo}, the aging of the individual is simulated with the dissipation given by the Lindblad dynamics in Eq. \eqref{lind}, and the mutations may be modelled with single qubit rotations as in Eq. \eqref{mut1}. Finally, the interaction among individuals given in Eq. \eqref{inter} involves a four-qubit operation, which can be decomposed in terms of the Toffoli gate, $U_{\rm CCNOT}$, by relabeling the levels encoding the quantum state. 
\begin{equation}
\{|4\rangle, |7\rangle \} \rightarrow \{|7\rangle, |8\rangle\}, \hspace{0.2 cm} \{|10\rangle, |13\rangle \} \rightarrow \{|15\rangle, |16\rangle\} \implies U_I \rightarrow \mathbb{1} \otimes U_{\rm CCNOT}
\end{equation}
 
\subsubsection*{Trapped Ions} In trapped ion devices, chains of ions are spatially confined by using time dependent electromagnetic fields. The physical system is described with a model that results in a set of internal (electronic) and collective (motional) quantized energy levels. Laser fields tuned in the frequencies of the desired transitions provide a good control of the system allowing for the implementation of several physical operations that can be translated into logical ones. In particular, we show the trapped-ion control Hamiltonian for a single spin and a single mode, 
\begin{equation}
\label{ion}
H= \hbar \Omega \sigma^{+} \left[ 1 + i \eta \left( a e^{-i \nu t}+ a^{\dag} e^{i \nu t} \right) \right]e^{i(\phi-\delta t)} + \textrm{H.c.}
\end{equation}
Therefore, trapped ions can be used as quantum information processing experimental platforms. Here, $\Omega$ is the Rabi frequency, $\sigma^{+}$ the spin raising operator, $\eta$ the Lamb-Dicke parameter, $a$ and $a^{\dag}$ the motional annihilation and creation operators, $\nu$ the trap frequency,  $\phi$ the initial phase of the laser field and $\delta$ the detuning between the laser and qubit frequencies.

In our proposal, each quantum living unit, composed of two qubits, can be encoded either in four metastable levels of a single ion or two levels of a pair of ions, and ancillary levels enable the readout of the state \cite{ion}. The $U_{\rm CNOT}$ can be implemented with the M\o lmer-S\o rensen gate and a sequence of single-qubit gates \cite{moso}. We rely on previous proposals \cite{opsy} for simulating the dissipative dynamics. Mutations as single qubit rotations could be done with controlled Rabi oscillations between the levels that encode the genotype. The interaction among individuals can be realized with the Toffoli gate, that has already been implemented in a trapped-ion setup \cite{toff}.

\subsubsection*{Superconducting Circuits} The superconducting circuits are designed with inductors, capacitors and Josephson junctions as building elements. The main properties of the circuits are the superconductivity that allows the flow of electrical current without energy dissipation, and the nonlinear separation of the quantized energy levels introduced by the Josephson junctions. The effective equation that describes the time evolution of the charge and phase, the degrees of freedom in the superconducting circuit, can be manipulated by rearranging the disposition of the circuit elements. The system can be controlled by coupling it with resonant photons, which allows for the implementation of several physical models, including the Jaynes-Cummings model, 
\begin{equation}
\label{scc}
H= \omega_r a^{\dag} a + \epsilon \sigma^{+}\sigma^{-} + g(a \sigma^{+} + a^{\dag} \sigma^{-}).
\end{equation}
Here $\omega_r$ is the photon frequency, $g$ is the coupling constant, and $\epsilon$ is the qubit frequency, which is encoded in the quantum excitations of the circuit. 

For the particular aspects of our proposal, the transmon is the most appropriate superconducting qubit, because of its long coherence time. The single and two qubit gates used for implementing self-replication and mutations can be realized with high fidelities~\cite{gamb}. The controlled dissipation necessary for encoding the evolution of the phenotype can be realized with current technology~\cite{prx}. The Toffoli gate performing the interaction processes among individuals is feasible in superconducting circuits \cite{wall}. 

\subsubsection*{Quantum Photonics} In quantum photonics devices, quantum information processing tasks are implemented with linear optical elements: beam splitters, phase shifters, single photon sources and photodetectors. The qubit is encoded in the coherent superposition of two modes in a photon. The variation in the refractive index, and the interaction between the modes are the physical mechanism introduced by the phase shifters and beam splitters respectively. These operations allow for the realization of deterministic single qubit gates, and are modeled with the following Hamiltonians,
\begin{equation}
H_{ps}=\phi a^{\dag}_{in} a_{in}, \qquad H_{bs}=\theta e^{i\phi} a^{\dag}_{in}b_{in} + \theta e^{-i \phi} a_{in} b^{\dag}_{in}
\end{equation}
Here, $a$ and $b$ are the modes in which the qubit is encoded, $\phi$ is the relative phase and $\theta$ the phase associated with the transmission amplitude. Moreover, two qubit gates are performed probabilistically, employing the Hong-Ou-Mandel effect as a computational resource.

These techniques provide a complete set of single and two qubit gates, that can be extended for particular logical operations, like the Toffoli gate~\cite{phot}. Therefore, the self-replication, mutations and interactions in our model can be realized in an experimental protocol with photons. Finally, a technique for the implementation of stochastic quantum walks~\cite{nmar} could be used to simulate the evolution of the phenotype.

\section*{Discussion}
We have developed a quantum information model for mimicking the behaviour of biological systems inspired by the laws of natural selection. Our protocol is hardly tractable with classical simulations, leaving many relevant questions coming from the classical models without answer in our quantum analogue. This justifies an experimental implementation of these ideas in a controllable quantum platform. Simultaneously, we have analyzed several figures of merit, which provide partial information about the quantum features of the model for small systems. Finally, we have studied the feasibility of the protocol in different physical systems, which enables the realization of artificial life in quantum technologies.

\section*{Acknowledgements}
The authors acknowledge inspiring discussions with O. Boada and Y. Omar, and support from Spanish MINECO FIS2012-36673-C03-02; Ram\'on y Cajal Grant RYC-2012-11391; UPV/EHU UFI 11/55 and EHUA14/04, Basque Government IT472-10 and BFI-2012- 322; PROMISCE and SCALEQIT EU projects.

\section*{Author Contribution}
U.A.-R. made the calculations while U.A.-R., M.S., L.L. and E.S. developed the protocol and wrote the manuscript.

\section*{Additional information}
The authors declare no competing financial interests.

\end{document}